\documentstyle[12pt,epic,eepic]{article}

\newtheorem{theorem}{Theorem}[section]
\newtheorem{lemma}[theorem]{Lemma}
\newtheorem{corollary}[theorem]{Corollary}

\newtheorem{definition}{Definition}
 
\newenvironment{proof}{\par\noindent {\em Proof. }}{\myendproof}

\newcommand{\algboxB}[3]{
\begin{center}
\fbox{\hspace{#2}\parbox{#1}{\vspace{#2}#3
\parbox{#1}{\vspace{#2}}}\hspace{#2}}%
\end{center}}

\newenvironment{tabAlgorithm}[2]{
\setcounter{algorithmLine}{1}
\samepage
\begin{tabbing}
999\=\kill
#1 \ \ --- \ \ \parbox{3.8in}{\it #2}
}{
\end{tabbing}
}
\newcounter{algorithmLine}
\newcommand{\algline}{\\\thealgorithmLine\hfil\>\stepcounter{algorithmLine}}
 \newcommand{\algnono}{\\ \>}

\date{}

\def\myendproof{\hfill{\vbox{\hrule\hbox{%
   \vrule height1.3ex\hskip0.8ex\vrule}\hrule}}}

\newcommand{\C}[1]{\beta_{#1}}

\newcommand{\fourth}{\frac{1}{4}}

\newcommand{\fig}[3] %usage:\fig{file}{label}{caption}
{\begin{figure}[hbtp]
 \begin{center}
 \input{#1}
 \end{center}
 \caption{#3}
 \label{#2}
 \end{figure}
}

\newcommand{\topfig}[3] %usage:\topfig{file}{label}{caption}
{\begin{figure}[tp]
 \begin{center}
 \input{#1}
 \end{center}
 \caption{#3}
 \label{#2}
 \end{figure}
}

\newcommand{\TreeAlgorithm}[1]{
\begin{figure}[hbp]
\algboxB{4.8in}{0.1in}{
\begin{tabAlgorithm}{{\sc Route-Tree}$(S)$}%
{Find a routing tree for $S$.}
\algline {\bf If} $|S| = 1$ {\bf then Return} $S$ as a tree on a
single vertex.
\algline For each $v\in S$, fix its weight $U(v)$ to be $W(\{v\},\bar S)$.
\algnono   Let the sum of the weights of the vertices in $S$ be $U_S$.
\algline {\bf If} \= for any vertex $v$, $U(v) \ge U_S/2$ and $U_S\ne 0$ {\bf then}
\algline        \> {\sc Route-Tree}$(S\setminus\{v\})$
\algline        \> Create a new tree $T$ by attaching the above tree
                        and $v$ as the children \\\>\>
                        of a new root $r$. {\bf Return} $T$.
\algline Find an approximate minimum-weight $\fourth$-balanced
         separator for the \\\> subgraph induced by $S$ in $G$ 
        (if $U_S=0$, find an unweighted balanced \\\> separator). Let this
         break $S$ into pieces $S_1$ and $S_2$.
\algline {\sc Route-Tree}$(S_1)$
\algline {\sc Route-Tree}$(S_2)$
\algline Create a new tree $T$ by attaching the two trees generated
         above as the \\\> children of a new root vertex. {\bf Return} $T$.
\end{tabAlgorithm}
\vglue-\baselineskip
}
\vglue-\baselineskip
\caption{Approximation Algorithm to Find a Routing Tree}
\label{routing-tree-algorithm}
\end{figure}
}

\title{Designing Multi-Commodity Flow Trees}

\author{Samir Khuller 
\thanks{Department of Computer Science and Institute for
Advanced Computer Studies, University of
Maryland, College Park, MD~20742.
Research currently supported by NSF Research Initiation Award CCR-9307462.
 E-mail : {\tt samir@cs.umd.edu}.}
\and Balaji Raghavachari
\thanks{Department of Computer Science, The University of Texas at Dallas,
Box 830688, Richardson, TX 75083-0688. E-mail : {\tt rbk@utdallas.edu}. Part of
this work was done while this author was visiting UMIACS.}
\and Neal Young
\thanks{Institute for Advanced Computer Studies, University of
Maryland, College Park, MD~20742. E-mail : {\tt young@umiacs.umd.edu}.
Research supported in part by NSF grants CCR-8906949 and CCR-9111348.}
}
\date{ }
\begin{document}

%\begin{titlepage}
\thispagestyle{empty}
\maketitle
\thispagestyle{empty}
\begin{abstract}
The traditional multi-commodity flow problem 
assumes a given flow network in which multiple
commodities are to be maximally routed
in response to given demands.
This paper considers the multi-commodity flow network-design problem:
given a set of multi-commodity flow demands, 
find a network subject to certain constraints 
such that the commodities can be maximally routed.

This paper focuses on the case when the network is required to be a tree.  
The main result is an approximation algorithm 
for the case when the tree is required to be of constant degree.
The algorithm  reduces the problem to 
the minimum-weight balanced-separator problem;
the performance guarantee of the algorithm
is within a factor of 4 
of the performance guarantee of the balanced-separator procedure.
If Leighton and Rao's balanced-separator procedure is used,
the performance guarantee is $O(\log n)$.
%js
%rbk
This improves the $O(\log^2 n)$ approximation factor obtained
by a direct application of the balanced-separator method.
\end{abstract}

%\end{titlepage}

\section{Introduction}

Let a graph $G = (V, E)$ represent multicommodity flow demands:  
the weight of each edge $e=\{a,b\}$ 
represents the demand of a distinct commodity 
to be transported between the sites $a$ and $b$.  
Our goal is to design a network,
in which the vertices of $G$ will be embedded,
and to route the commodities in the network.
The maximum capacity edge of the network should be low
in comparison to the best possible in {\em any} network meeting
the required constraints.
For example, the weight of each edge could 
denote the expected rate of phone calls between  two sites.  
The problem is to design a network in which calls can be routed
minimizing the maximum bandwidth required; 
the cost of building 
the network increases with the required bandwidth.

We consider the case when the network is required to be a tree,
called the {\em tree congestion problem}.
Given a tree in which the vertices of $G$ are embedded,
the load on an edge $e$ is defined as follows: 
delete $e$ from $T$. This breaks $T$ into
two connected components. If $S$ is the set of vertices from $G$ in
one of the connected components, then $load(e)$ is equal to
\[ 
W(S,\bar S) = \sum_{(x,y) \in E, x \in S, y \in \bar S} w(x,y). 
\] 
In other words, the demand of each edge $e=\{a,b\}$ in $G$, maps to the unique
path in $T$ from $a$ to $b$, and loads each edge on the path.  
The load of a single edge is the sum of the demands that load 
this edge.

In this paper we study two different versions of this problem.

\subsection{Routing Tree Problem}

The following problem was proposed and studied by 
Seymour and Thomas~{\cite{ST}}.

\medskip

%rbk removed the citation from the definition. We already mention it above.
\begin{definition}  %\cite{ST}
A tree $T$ is called a {\em routing tree} if it  satisfies the following
conditions:
\begin{itemize}
\item The leaves of $T$ correspond to vertices of $G$.
\item Each internal vertex has degree 3.
\end{itemize}
The congestion of $T$ is the maximum load of any edge of
$T$. The congestion of $G$, denoted by $\C{G}$, is defined to be the minimum 
congestion over all routing trees $T$ of $G$.
\end{definition}

We would like to find a routing tree $T$ with minimum congestion
(that achieves $\C{G}$).

Seymour and Thomas showed that this problem is
NP-hard by showing that graph bisection can be reduced to this
problem.  They also showed that in the special case when $G$ is
planar, the problem can be solved optimally in polynomial time.

We provide a polynomial time approximation algorithm for the
congestion problem when $G$ is an arbitrary graph. Our algorithm
computes a routing tree $T$  whose congestion is within an
$O(\log n)$ factor from the optimal congestion (Section \ref{rbksec}).
The algorithm extends to the case when the routing tree is allowed to have
vertices of higher degree.

\subsection{Congestion Tree Problem}
We also study the case when $T$ is required to be a spanning tree of 
a given feasibility graph $F$. 
We show that the problem is NP-complete (Section~\ref{gen-cong-sec}).
In the special case when $F$ is complete, we show that an optimal
solution can be computed in polynomial time.
%rbk I think that the referee is correct in objecting to this. I
%suggest we get rid of the stuff about GH trees. Talking about the GH
%tree of a demand graph being a subgraph of an entirely independent
%feasibility graph (supply graph) is not meaningful.
\iffalse
\footnote{We 
actually show that if  the Gomory-Hu cut tree $T_{GH}$ 
of $G$ \protect\cite{GH,Gu} is a subgraph of $F$
then $T_{GH}$ is an optimal solution.}.
We conjecture that  using ideas similar to the ones
used to solve the routing tree problem, one can design an $O(\log n)$
approximation algorithm for the congestion tree problem.
%js replaced scheme by algorithm
\fi

\subsection{Main Ideas}

Our algorithm is a simple divide-and-conquer algorithm that uses the
Leighton-Rao~\cite{LR} balanced separator algorithm to split the graph. 
By a naive application of the LR algorithm, one obtains an 
$O(\log^2 n)$ approximation factor. Our main contribution is to show that
by a subtle application of LR, one can actually obtain an $O(\log n)$ 
approximation factor.
%rbk replaced suspect with believe and got rid of "actually"
%We suspect that this kind of an application of LR
%will actually be useful for other problems as well 
We believe that this kind of an application of LR
%samir
%will prove to be useful for other problems as well 
will prove to be useful in obtaining better approximation
ratios for other problems as well.
%rbk  Two factors two close to each other.
%(in improving approximation factors by a factor of $\log n$).
%(and improve the approximation ratio by a factor of $\log n$).
%rbk The above line has the plural-singular problem. What is the right
%form? We talk about applying it to other problems, in which case we
%should improve their approximation factors (plural).

%rbk This section needs to be restructured properly. Right now it
%appears as if LR addresses the case when vertices are weighted.
% I have copied the section as it used to be and put it after the end
% of the paper.
\section{Preliminaries}

%rbk rewrote the following sentence. If you want the original form,
%feel free to put it back.
%A {\em cut} in a graph $G$ is a set of edges which separate $G$ into
%two pieces $S$ and $\bar S=V\setminus S$. 
A {\em cut} in a graph $G$ is a set of edges whose removal separates $G$ into
two disconnected pieces $S$ and $\bar S=V\setminus S$. 
A cut can be represented by
the vertex set $S$. The {\em weight} of a cut $S$, denoted by
$W(S,\bar S)$, is the sum of the weights of those edges which have one
endpoint in $S$ and one endpoint in $\bar S$. 
We use $W(v)$ to refer to the sum of the weights of the edges
incident to $v$. 
%rbk switched b and n below.
A cut $S$ is {\em $b$-balanced} if $b\cdot n \le |S| \le (1-b)\cdot n$. The
definition is extended to the case when vertices are weighted as
follows. Let $U$ be a non-negative weight function on the vertices and
let $U(S)$ be the sum of the weights of all the vertices in $S$. A cut $S$ is
$b$-balanced if

\[
b\cdot U(V) \le  U(S) \le (1-b) \cdot U(V)
\]

\begin{definition}
%N A {\em $\lambda$-approximate minimum $b$-bisector} is a $b$-balanced
   For $b \le 1/3$,
   a {\em $\lambda$-approximate minimum $b$-bisector} is a $b$-balanced
cut whose weight is at most $\lambda$ times the weight of a
%N minimum-weight $\frac{1}{3}$-balanced cut, for some constant
   minimum-weight $\frac{1}{3}$-balanced cut.
%N $b\le\frac{1}{3}$.
\end{definition}

The following result was proved by Leighton and
Rao (\cite{LR}, Section 1.4).

%rbk The following theorems are not verbatim from the papers. I
%removed their citations from the statement of the theorems.
\begin{theorem}  %[\cite{LR}]
%N Let $G$ be a graph with non-negative weights on the edges and
   Let $G$ be a graph with non-negative weights on the edges
%N let the vertices of $G$ have uniform weights (unweighted).
   (without vertex weights).
It is possible to compute an $O(\log n)$-approximate
minimum $\fourth$-bisector of $G$ in polynomial time.
\end{theorem}

The above theorem was extended to the case when vertices are given
non-negative weights by Tragoudas~\cite{Tr}.

\begin{theorem} \label{weighted-separator-theorem}  %[\cite{Tr}]
Let $G$ be a graph with non-negative weights on the edges and vertices.
It is possible to compute an $O(\log n)$-approximate
minimum $\fourth$-bisector of $G$ in polynomial time.
\end{theorem}

%rbk This should be replaced by a reference to Tragoudas.
\iffalse
The above theorem can be extended to the case when vertices are given
non-negative weights~\cite{Rao,Tar}.
\fi

\begin{definition}
Let $T$ be a tree and let $u$ be a vertex of degree two in $T$. Let
$v$ and $w$ be the neighbors of $u$. The following operation is said
to {\em short-cut} $u$ in $T$ -- delete $u$ from $T$ and add the edge
%N $\{v,w\}$. Short-cutting $T$ implies the deletion of all vertices of
   $\{v,w\}$. To short-cut $T$ is to delete all vertices of
degree two by short-cutting them in arbitrary order.
\end{definition}

\section{Routing Tree Problem}
\label{rbksec}

$W(v)$ corresponds to the total weight
between $v$ and other vertices and is called the load of a vertex.
Note that the load of any vertex $v$ is a lower bound on $\C{G}$,
because the edge incident to the leaf corresponding to $v$ in any
routing tree has to handle this load.

\begin{lemma} \label{leaf-lower-bound}
For any vertex $v$,  $W(v) \le \C{G}$.
\end{lemma}

Given a procedure to compute a $\lambda$-approximate minimum
$b$-bisector, our algorithm finds a routing tree whose congestion is
at most $\lambda/b$ times the optimal congestion.

\subsection{Lower Bounds}
We show two ways of finding lower bounds on the weight of the optimal
solution. First, we show that the weight of a minimum-weight balanced
separator is a lower bound on $\C{G}$. 
Second, we show that the optimal solution for the problem in a
subgraph $G^\prime$ induced by an arbitrary set of vertices
$V^\prime\subset V$ is a lower bound on the optimal solution of $G$.
This implies that an optimal solution to a sub-problem costs no more than
any feasible solution to the whole problem.

\begin{lemma} \label{separator-lemma}
Let $G=(V,E)$ be a graph with non-negative weights on the
edges.  Suppose we are given a non-negative weight function $U(v)$ on the
vertices. Let the weight of each vertex be at most one-half of the
total weight of all the vertices. Let $Q$ be the weight of a 
minimum-weight $b$-balanced separator of $G$ for any $b\le1/3$. 
Then $Q \le \C{G}$.
\end{lemma}

\begin{proof}
Let $T$ be a routing tree with congestion $\C{G}$.  Each edge $e$ of
$T$ naturally induces a cut in $G$ as follows: delete $e$ from $T$ to
obtain subtrees $T_1$ and $T_2$. Let $S_e$ be the set of vertices in
$G$ that are leaves of $T_1$ (this yields a cut in $G$).  Clearly,
$W(S_e,\overline{S_e})$ is the congestion on edge $e$ and hence
$W(S_e, \overline{S_e}) \leq \C{G}$.  Since $T$ is a tree of degree
three, and by the assumption on the weights of  vertices, 
it contains at least one edge $e'$ which yields a
$b$-balanced separator. Since $Q$ is the minimum
$b$-balanced separator of $G$ we have $Q \leq W(S_{e'},
\overline{S_{e'}}) \leq \C{G}$.
\end{proof}

\begin{lemma} \label{subgraph-lemma}
Let $G=(V,E)$ be a graph. Let $H$ be a subgraph of $G$.
Then $\C{H} \le \C{G}$.
\end{lemma}

\begin{proof}
%js made the proof a little shorter...
%N Let $T$ be a routing tree with congestion $\C{G}$. It is easy to generate a
   Let $T$ be a routing tree with congestion $\C{G}$. Generate a
%N routing tree $T_H$ for $H$ from $T$ such that the load of any edge in
   routing tree $T_H$ for $H$ from $T$ as follows.
%N $T_H$ is at most the load of some edge in $T$.
%We generate the tree
%$T_H$ from $T$ as follows.  
Let $V_H$ be the vertex set of $H$. Mark
the leaves of $T$ corresponding to $V_H$. Repeatedly delete the
unmarked leaves of $T$ until it has no unmarked leaves. Delete all
vertices of degree two by short-cutting the tree, thus yielding $T_H$.
%The tree that we generate has $V_H$ as its leaves and all its internal
%vertices have degree three. 
%N This is a routing tree for $H$. Cuts
%N in $T_H$ can be associated with corresponding cuts in $T$ and hence
%N the load on any edge in $T_H$ is at most  the load of its
%N corresponding edge in $T$.
   It is easily verified that $T_H$ is a routing tree for $H$
   with congestion bounded by $\C{G}$.
\end{proof}

\subsection{The Routing Tree Algorithm}
%N
  \paragraph{Discussion.}
%N Our basic approach is to subdivide the graph into pieces which are
   Our basic approach is to subdivide the graph into pieces that  are
smaller by a constant fraction using an approximately minimum bisector.
Since computing a minimum-weight
balanced separator is also NP-hard, we use approximation algorithms
%rbk added Tragoudas
designed by Leighton and Rao~\cite{LR} and Tragoudas~\cite{Tr} for computing
approximately minimum-weight balanced separators (or approximate
minimum bisectors). 
The solutions for the pieces are obtained recursively. All internal
vertices of the solution tree have degree three except for the root.
The two trees are glued together by creating a new root and making the
%N roots of the pieces as the children of the new root. If implemented
   roots of the pieces    the children of the new root. If implemented
naively, this procedure leads to an $O(\log^2n)$ factor approximation.
Using balancing techniques, we improve the performance ratio to
$O(\log n)$.

Suppose $S$, a subset of the vertices representing a subproblem,
is split into two pieces $S_1$ and $S_2$ using an approximate
bisector. When the problem is solved recursively on the two pieces,
the main obstacle to obtaining an $O(\log n)$ approximation is the
following. 
In the worst case, it is possible that most of the load
corresponding to $W(S,\bar S)$ may fall on $S_1$ or $S_2$. 
If this happens repeatedly, an edge can be overloaded proportionally to its 
depth in the tree.
{\em To avoid this, it is
necessary to partition the demand from $\bar S$ roughly equally among the
pieces $S_1$ and $S_2$.} The following idea solves the
problem and leads to an $O(\log n)$ approximate solution.  Suppose we
define a weight $U(v)$ for each vertex $v$ in $S$ according to the amount
of demand from $v$ to the set $\bar S$. Now when we split $S$, we use a
%N cut that splits the vertices of $S$ into     sets of roughly equal weights.
   cut that splits the vertices of $S$ into two sets of roughly equal 
   {\em weight}.
Lemma~\ref{separator-lemma} guarantees that the minimum value of
such a cut is a lower bound on $\C{S}$, which is a lower bound on
$\C{G}$ by Lemma~\ref{subgraph-lemma}.
We illustrate the recursive step of the 
algorithm by an example in Fig.~\ref{ex1}.

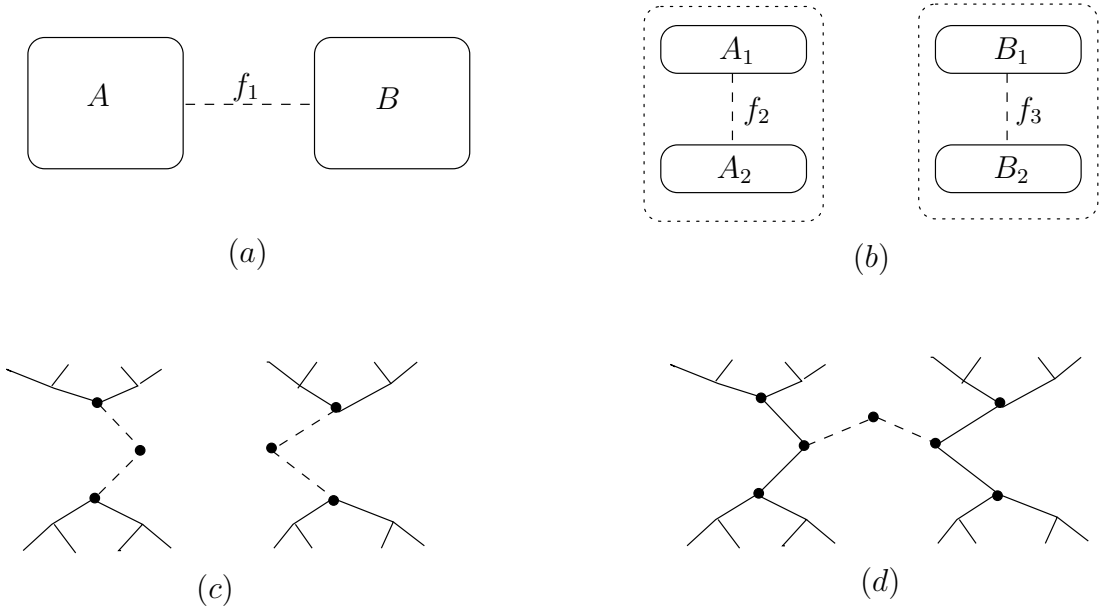
\topfig{ipl-ex1}{ex1}{Example to illustrate algorithm.}

The algorithm first splits graph $G$ into $A,B$ by using an
%N approximate bisector. Each vertex in $A$ is assigned a weight
   approximate bisector (without weighting the vertices).
   Each vertex in $A$ is then assigned a weight
equal to the total demand it has to vertices in $\bar A$. Similarly
vertices in $B$ are assigned weights corresponding to their demands
  from $\bar B$. The algorithm now recursively splits $A$ and $B$ by
%N approximate bisectors.
   approximate bisectors with respect to the vertex weights.
%N The weight of each vertex in $A_1$ is now increased
%N by its demand to vertices in $A_2$ (similarly for sets $A_2,B_1,B_2$).
The problem is solved recursively on each piece. These recursive calls
%N
   weight vertices similarly and
return with respective trees as solutions for the pieces $A$ and $B$
as shown. By adding new edges and a new root vertex, the solution for
the entire graph is obtained.

\TreeAlgorithm

The algorithm given in Fig.~\ref{routing-tree-algorithm} implements
the above ideas. 
%js: added the foll para
The procedure{\sc Route-Tree}$(S)$  takes a subset of vertices $S$,
and returns a routing tree for the graph induced by the vertices in $S$.
This routing tree will either be a singleton vertex, or a tree
in which each vertex has degree one or  three, except for the root that
 has degree two.
The routing tree is computed in a way so as to approximately ``divide''
the demand from the vertices in $S$ to the vertices in $V-S$.

%N
  \paragraph{Analysis.}
Given a graph $G$, {\sc Route-Tree}$(V)$ returns
a routing tree for $G$. To make sure that the root of the tree has
degree three, we can discard the root by short-cutting it.

%rbk where the algorithm used to be

Let the algorithm use a $\lambda$-approximate minimum
$\fourth$-bisector in Line 6. If Leighton and Rao's \cite{LR} balanced
separator algorithm is used, $\lambda=O(\log n)$. The following
theorem shows that the load of any edge is at most $4\lambda$ times
the optimal congestion. We use induction to prove that our
load-balancing technique splits the load properly.

\begin{theorem}[Performance] \label{rt-perf-theorem}
The algorithm in Fig.~\ref{routing-tree-algorithm} finds a routing
tree $T$ for $G$ such that $\C{T} \le 4\lambda\C{G}$.
\end{theorem}

\begin{proof}
The proof proceeds by induction on the level of recursion. In the
%N first call of {\sc Route-Tree}, $G$ is split into two pieces $S$ and
   first call of {\sc Route-Tree}, the algorithm splits $G$ 
   into two pieces $S$ and
%N $\bar S$ using an approximate bisector. We then find  routing trees for $S$
   $\bar S$ using an approximate bisector. It then finds routing trees for $S$
%N and $\bar S$ and connect  the two roots with an edge $e$. The load
   and $\bar S$ and connects the two roots with an edge $e$. The load
on $e$ is $W(S,\bar S)$. By Lemma~\ref{separator-lemma},
the weight of a minimum-weight balanced separator is a lower bound on
$\C{G}$. The weight of the separator the algorithm uses is guaranteed
to be at most $\lambda$ times the weight of an optimal separator. Hence
the load on edge $e$ is at most $\lambda\C{G}$. This satisfies the
induction hypothesis.

\fig{ipl-ex2}{ex2}{Inductive proof.}

For the induction step, let us consider the case when we take a set $S$ and
split it into two pieces $S_1$ and $S_2$ (see Fig.~\ref{ex2}).  
Let $L$ be the load on the edge connecting the tree for $S$
to its parent.
Similarly, let $L_i$ $(i=1,2)$ be the load on the edge connecting the tree 
for $S_i$ to its parent.
Inductively, $L \leq 4 \lambda \C{G}$. 
We show that each $L_i \leq 4 \lambda \C{G}$. 

Let $U$ be the weight function defined by the algorithm in this recursive call.
Note that $L = U(S) = W(S,\bar S)$ and
$L_i = W(S_i, \bar S_i) = W(S_i, \bar S) + W(S_1,S_2)$.
Also observe that $U(S_i) = W(S_i,\bar S)$.

{\em Case 1:} 
If there is some vertex $v$ in $S$ whose
weight $U(v)$ is more than $U(S)/2$, then we split $S$ as $S_1=\{v\}$ and
$S_2 = S\setminus\{v\}$. 
Since $L_i = U(S_i) + W(S_1,S_2)$ and 
$U(S_1) > U(S)/2 > U(S_2)$ it follows that $L_1 > L_2$.  
This is because $U(S)$ is the sum of $U(S_1)$ and $U(S_2)$.
It remains only to bound $L_1$.
The demand from $v$, $W(v)$, is a lower bound on the
congestion (by Lemma~\ref{leaf-lower-bound}) and 
therefore $\C{G} \ge W(v)= L_1$. Hence both $L_1$ and $L_2$
satisfy the induction hypothesis.

{\em Case 2:} Otherwise, the algorithm distributed $U(S)$ into the weights
of the vertices of $S$ and then used a $\lambda$-approximate 
$\fourth$-bisector of $S$.
By the induction hypothesis, the edge from the subtree of $S$ to its 
parent has a load $L$ $(= U(S))$ of at most $4\lambda\C{G}$. 

Since $W(S_i, \bar S) = U(S_i) \leq \frac{3}{4} U(S)$ and
$W(S_1,S_2) \leq \lambda\C{G}$ (by Lemmas~\ref{separator-lemma} and 
\ref{subgraph-lemma}) we have:
\[ L_i = W(S_i, \bar S) + W(S_1,S_2) \leq 3\lambda\C{G} + \lambda\C{G}.\]
\end{proof}

\begin{theorem}%js [Running Time]
The routing tree algorithm in Fig.~\ref{routing-tree-algorithm} runs
in polynomial time. \myendproof
\end{theorem}

\begin{corollary}
The algorithm in Fig.~\ref{routing-tree-algorithm} finds in polynomial time a
routing tree $T$ for $G$ such that $\C{T} = O(\log n)\C{G}$.
\end{corollary}

\noindent
{\bf Note:}
Our algorithm also handles the case when vertices of $G$ are allowed to be
internal vertices of the output tree. Lemmas~\ref{separator-lemma} and
\ref{subgraph-lemma} are valid in this case also. The lower bound in
Lemma~\ref{leaf-lower-bound} weakens by a factor of 3. This lower bound is not
critical to the performance ratio, so the performance ratio of the
algorithm is unchanged. 

Our algorithm can be generalized to find routing trees when every 
internal vertex may have degree up to $k$, for any $k\geq 3$. We
obtain the same $O(\log n)$ approximation factor, independent of $k$.
An algorithm obtaining an approximation factor of $n/k$ is straightforward 
and is useful as $k$ approaches $n$.

\section{General Congestion Problem}
\label{gen-cong-sec}
%rbk Reducing the size of the following section. I think that no
%proofs are really necessary, but I am not removing the NPC proof.
%\subsection{NP-Completeness}
In this section we show that the following problem is NP-complete.
%N
\iffalse
Given a graph $G = (V, E)$ representing a demand network.
Each edge $e = \{a,b\}$ has a nonnegative weight 
$w(e)$ that represents the demand between the sites $a$ and $b$.  
We are also given a feasibility graph $F$ and an integer $D$. 
\fi
  The input to the problem is a demand network $G=(V,E)$,
  a ``feasibility network'' $F=(V,E')$, and an integer $D$.
  Each edge $e = \{a,b\}$ of $G$ has a nonnegative weight 
  $w(e)$ that represents the demand between the sites $a$ and $b$.  
The problem is to find a tree $T$ that is a subgraph of $F$, 
such that when the demands of the edges in $G$ are mapped to the tree $T$
the congestion on each edge is at most $D$.

The reduction is done from the $k$ Edge-Disjoint Paths Problem, known to 
be NP-Complete~\cite{GJ}. 
%We show that
%this problem is NP-complete by a reduction from the 2-Commodity Flow Problem.

\noindent
{\bf $k$ Edge-Disjoint Paths Problem:}
Given an undirected graph $H = (V,E)$, and sets 
$S = \{ s_1, s_2, \ldots, s_k \}$ and $T = \{ t_1, t_2, \ldots, t_k \}$
are there $k$ mutually edge-disjoint paths $P_1, P_2, \ldots, P_k$
such that $P_i$ connects $s_i$ with $t_i$ ?

It is easy to see that this problem can be reduced to the general tree
congestion problem. For the reduction we construct $F$ from $H$. 
For each vertex $u \in V$, if $u$ has degree $d(u)$, we
create a clique on $d(u)$ vertices, $u_1, u_2, ..., u_{d(u)}$. For
each edge from $v$ to $w$ we introduce an edge from $v_i$ to $w_j$
where these are {\em distinct} vertices (not shared with any other edges).
(Informally, each vertex is ``exploded'' into a clique, and the edges incident
on the vertex are made incident on distinct clique vertices.)  
The demand graph
$G$ has edges between $s_i$ and $t_i$ (for all $i$). If there is a solution
to the disjoint paths problem, clearly that yields a congestion tree with
bandwidth one. The set of paths $P_i$ can form cycles, but these cycles
can be ``pried'' apart in $F$ since we replaced each vertex with a clique. 
These can now be connected to form a congestion tree with bandwidth one.

If there is a solution to the congestion tree problem it is clear that
this yields a solution to the edge-disjoint paths problem (the demand
edge from $s_i$ to $s_j$ gets mapped to a path in the tree and causes
a load of one on each edge). Since the bandwidth is restricted to one, no
other path can use the same edge (even when we go from $F$ to $H$).   

%rbk added the following theorem.
\begin{theorem}
The general congestion problem is NP-complete.
\end{theorem}

%rbk Removing the following subsection and replacing it with a smaller
%paragraph.
\iffalse
\subsection{Polynomially Solvable Case}

In this section we show that when $T_{GH} \subseteq F$ 
(the feasibility graph contains the Gomory-Hu cut tree) we can
solve the congestion problem optimally. (This is certainly the case
when $F$ is a complete graph.)

Given the demand graph $G$, we compute the Gomory-Hu cut tree $T_{GH}$
\cite{GH,Gu}. This is the tree that is used to route the calls. 
This yields an optimal solution for the following reason: consider any
edge $e=\{s,t\}$ with load $L(e)$. $T_{GH}$ has the property that
$L(e)$ is the value of the $s$-$t$ min cut. Clearly any $s$-$t$ min
cut is a lower bound on the optimal congestion.

\begin{theorem}
$T_{GH}$ is an optimal solution to the congestion  problem.
\end{theorem}
\fi

An interesting open problem is to design approximation algorithms with
nontrivial approximation factors for designing routing trees where the
feasibility graph $F$ is given in the input.  In the special case
when $F$ is complete, it is easy to show that an optimal routing
tree can be computed. In this case each edge of the routing tree is
made to handle a load that is equal to the minimum cut in $G$
separating two of its vertices. This follows from the result of
Gomory and Hu~\cite{GH}, who showed how to construct a tree which
encodes {\em all} min-cuts in a graph. 
Gusfield~\cite{Gu} gave an algorithm to compute such trees efficiently.

\begin{theorem}
If $F$ is the complete graph, the problem of designing a
routing tree with minimum congestion for an arbitrary
demand graph $G$ can be solved in polynomial time.
\end{theorem}
 
%\section*{Acknowledgments}
%Do we want to ack anyone ?

\end{document}